\documentclass{article}
\usepackage{spconf,amsmath,graphicx,hyperref}
\usepackage{amsthm,amsfonts,amssymb}
\usepackage{multirow}
\usepackage{pifont}
\usepackage{colortbl}
\usepackage[ruled,linesnumbered,vlined]{algorithm2e}
\usepackage{xcolor}
\usepackage{booktabs}
\usepackage{circledsteps}
\usepackage[normalem]{ulem}

\definecolor{blue1}{rgb}{0.21,0.49,0.74}
\definecolor{red1}{rgb}{1,0,0}
\definecolor{green1}{rgb}{0.384,0.596,0.243}

\title{Physics-Aware Novel-View Acoustic Synthesis with Vision-Language Priors and 3D Acoustic Environment Modeling}

\name{
\shortstack{
    Congyi Fan$^{1}$,
    Jian Guan$^{1, *}$\thanks{*Corresponding author: j.guan@hrbeu.edu.cn},
    Youtian Lin$^{2}$,
    Dongli Xu$^{3}$,
    Tong Ye$^{1}$,\\
    Qiaoxi Zhu$^{4}$,
    Pengming Feng$^{5}$,
    Wenwu Wang$^{6}$
    }
}
\address{
    $^1$ Group of Intelligent Signal Processing, Harbin Engineering University, Harbin, China \\
    $^2$ School of Intelligence Science and Technology, Nanjing University, Suzhou, China \\
    $^3$ Processing Speech and Images, KU Leuven, Leuven, Belgium \\
    $^4$ Acoustics Lab, University of Technology Sydney, Ultimo, Australia \\
    $^5$ State Key Laboratory of Space Information System and Integrated Application, Beijing, China \\
    $^6$ Centre for Vision Speech and Signal Processing, University of Surrey, Guildford, UK 
 }

\begin{document}

\maketitle

\begin{abstract}

Spatial audio is essential for immersive experiences, yet novel-view acoustic synthesis (NVAS) remains challenging due to complex physical phenomena such as reflection, diffraction, and material absorption. Existing methods based on single-view or panoramic inputs improve spatial fidelity but fail to capture global geometry and semantic cues such as object layout and material properties. To address this, we propose Phys-NVAS, the first physics-aware NVAS framework that integrates spatial geometry modeling with vision–language semantic priors. A global 3D acoustic environment is reconstructed from multi-view images and depth maps to estimate room size and shape, enhancing spatial awareness of sound propagation. Meanwhile, a vision–language model extracts physics-aware priors of objects, layouts, and materials, capturing absorption and reflection beyond geometry. An acoustic feature fusion adapter unifies these cues into a physics-aware representation for binaural generation. Experiments on RWAVS demonstrate that Phys-NVAS yields binaural audio with improved realism and physical consistency.

\end{abstract}


\section{Introduction}
\label{sec:intro}

Spatial audio, conveying sound position, direction, and distance in 3D space, is essential for immersive applications such as augmented/virtual reality (AR/VR), gaming, and interactive media~\cite{gigante1993virtual,carmigniani2011augmented,lu2025deep}. A fundamental task is novel-view acoustic synthesis (NVAS), which generates binaural audio for arbitrary listener positions given a mono input and scene observations~\cite{liang2023av}. Realistic NVAS requires accurate modeling of sound propagation, governed by direct sound, early reflections, and reverberation, all influenced by scene geometry, object layout, and materials~\cite{maluski2004effect}.

The main challenge is that these factors jointly produce complex effects such as reflection, diffraction, and absorption; without explicitly modeling them, synthesized audio lacks spatial realism and physical consistency. Recent works thus leverage visual cues, motivated by the intuition that a scene’s appearance provides critical information for acoustics \cite{liang2023av,gao2024soaf,bhosale2024av,baek2025av,chen2025soundvista}.

\begin{figure}[!t]
    \centering
    \includegraphics[width=.95\linewidth]{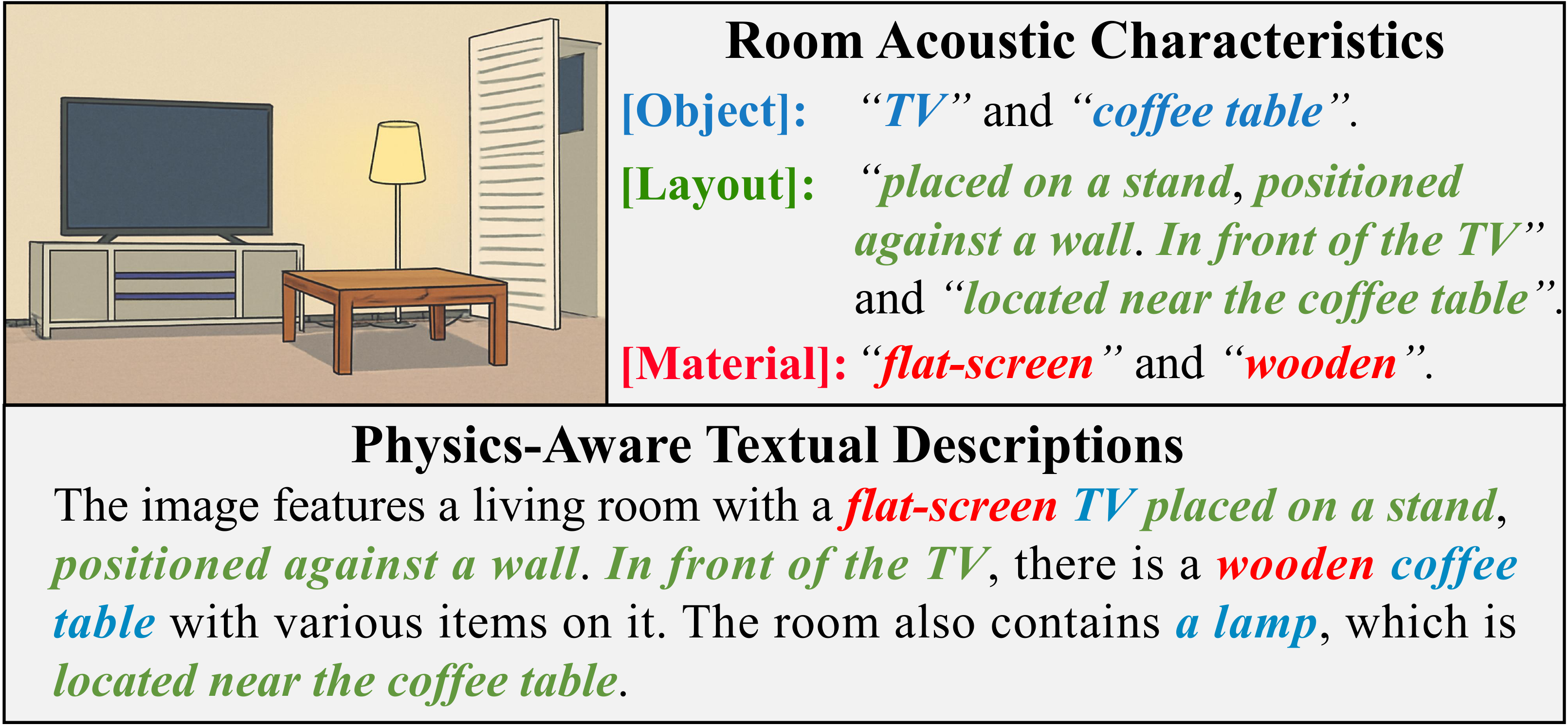}
    \caption{Illustration of scene semantics influencing acoustics. Different materials (e.g., \textit{wooden}, \textit{flat-screen}) affect absorption, while objects and their layouts (e.g., \textit{In front of the TV, there is a wooden coffee table}) modify reflection paths. These semantic cues are often ignored in existing NVAS methods, leading to physically inconsistent audio.}
    \label{fig:1}
    \vspace{-13pt}
\end{figure}
\begin{figure*}[!h]
    \centering
    \includegraphics[width=.93\linewidth]{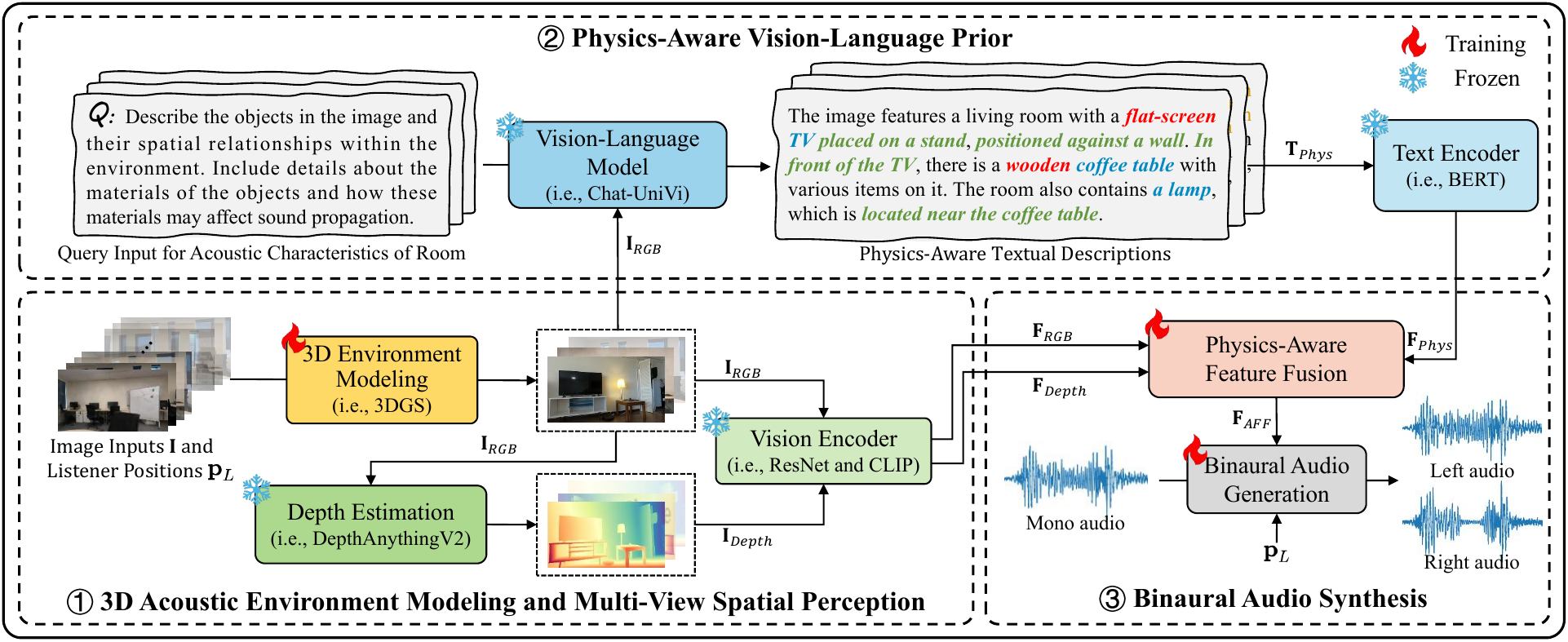}
   \caption{Overview of the proposed physics-aware NVAS framework. 3D acoustic environment modeling  with 3DGS and depth estimation on multi-view images, enhancing spatial awareness by recovering room geometry and size. Physics-aware vision–language priors further enrich acoustic  modeling  with object, layout, and material cues that capture absorption and reflection effects. Finally, geometric and semantic features are fused into a unified physics-aware feature representation, enabling realistic and physically consistent binaural audio generation.}
    \label{fig:2}
    \vspace{-12pt}
\end{figure*}

AV-NeRF~\cite{liang2023av} first conditioned a neural acoustic field on single-view images and depth, establishing a strong baseline on the Real-World Audio-Visual Scene (RWAVS) dataset. However, its reliance on listener-centric single views limited spatial awareness, and it did not explicitly incorporate semantic cues such as objects, layouts, and materials. Subsequent methods further enriched vision priors. For example, SOAF~\cite{gao2024soaf} modeled occlusion, AV-GS~\cite{bhosale2024av} exploited 3D Gaussian Splatting (3DGS)~\cite{kerbl20233d} for geometric fidelity, AV-Surf~\cite{baek2025av} refined structural details with surface normals, and SoundVista~\cite{chen2025soundvista} introduced panoramic context. While these methods improved geometry and occlusion handling, \textbf{\textit{they overlooked how object layout and material properties affect absorption and reflection.}} For instance, geometry-based models cannot distinguish the dampening of carpet versus tile, while panoramic inputs are weak in object-level semantics. As illustrated in Fig.~\ref{fig:1}, materials such as \textit{``wooden"} or \textit{``flat-screen"}, and layouts such as \textit{``In front of the TV"}, directly alter acoustic behavior. Ignoring such factors often leads to physically inconsistent audio synthesis.

In this paper, we propose Phys-NVAS, the first NVAS framework to incorporate physics-aware vision-language priors into scene acoustic modeling. 
Phys-NVAS integrates 3D acoustic environment reconstruction with physics-aware vision semantics to achieve more realistic spatial audio synthesis. 
Specifically, a global 3D acoustic scene is reconstructed with 3D Gaussian Splatting (3DGS) and depth estimation, enabling multi-view spatial perception and providing explicit structural cues on room geometry and size that guide direct sound and early reflections. Meanwhile, a vision-language model is employed to extract physics-aware priors describing objects, layouts, and materials, capturing absorption and reflection effects beyond geometry alone. We further propose an acoustic feature fusion adapter that integrates these geometric and semantic cues into a unified physics-aware representation for binaural generation. 
Experiments demonstrate that combining spatial geometry with semantic information yields more realistic and physically consistent novel-view acoustic synthesis.

\section{Proposed Method}
\label{sec:2}

This section presents the proposed Phys-NVAS framework, illustrated in Fig.~\ref{fig:2}. First, 3D acoustic environment modeling with 3DGS and depth estimation recovers global scene geometry and size, enhancing spatial perception via multi-view inputs. Second, a vision-language model extracts physics-aware priors such as objects, layouts, and material properties, providing semantic cues for absorption and reflection beyond geometry. Finally, an acoustic feature fusion adapter unifies geometric and semantic features into a physics-aware representation, enabling realistic and physically consistent binaural synthesis.

\subsection{3D Acoustic Environment Modeling and Multi-View Spatial Perception}

\noindent \textbf{\textit{3D  Environment Modeling:}}
To capture the global geometry of the acoustic environment, we generate multi-view RGB images $\mathbf{I}_{RGB}$ from a few image inputs $\mathbf{I}$ and the listener position $\mathbf{p}_L \in \mathbb{R}^5$, with 3D Gaussian Splatting (3DGS)~\cite{kerbl20233d} as image generator $\mathcal{G}(\cdot)$: 
\begin{equation}
    \mathbf{I}_{RGB} = \mathcal{G}(\mathbf{I}, \mathbf{p}_L).
\end{equation}
We then apply a pre-trained vision encoder (i.e., ResNet-18~\cite{he2016deep} and  CLIP~\cite{radford2021learning}) to extract geometric features $\mathbf{F}_{RGB} \in \mathbb{R}^{M}$ from $\mathbf{I}_{RGB}$, where $M$ denotes the feature dimension. Leveraging multiple viewpoints, $\mathbf{F}_{RGB}$ encodes global shape information of the scene that is relevant to spatial sound propagation.

\noindent\textbf{\textit{Multi-View Depth Perception:}} 
While RGB features describe global structure, they lack explicit geometric cues such as distances between listener and objects. To address this, we apply a depth estimation model $\mathcal{D}(\cdot)$ (i.e., DepthAnythingV2~\cite{yang2024depth2}) to generate multi-view depth maps $\mathbf{I}_{Depth}$ as follows:
\begin{equation}
    \mathbf{I}_{Depth} = \mathcal{D}(\mathbf{I}_{RGB}).
\end{equation}
The depth maps are then encoded into structural features $\mathbf{F}_{Depth}\in \mathbb{R}^{M}$. With the estimated distances between the listener and surrounding objects, $\mathbf{F}_{Depth}$ provides explicit cues about room size and shape, thereby complementing $\mathbf{F}_{RGB}$ for modeling direct sound propagation and early reflections in the acoustic environment.

\subsection{Physics-Aware Vision-Language Priors}
While multi-view geometry provides global structural cues, it fails to capture semantic properties such as objects, layouts, and materials that critically affect sound absorption and reflection. To address this limitation, we introduce a physics-aware vision-language prior that extracts semantic descriptions from the rendered multi-view images.

Specifically, we employ a vision-language model (VLM), i.e., Chat-UniVi~\cite{jin2024chat}, to generate textual descriptions $\mathbf{T}_{Phys}$ containing object identities, spatial layout, and material attributes from $\mathbf{I}_{RGB}$. These physics-aware textual priors explicitly describe scene configurations relevant to acoustic modeling (e.g., ``\textit{a flat-screen TV}'', ``\textit{a wooden coffee table}'', and their relative positions). To obtain such descriptions, we provide the VLM with a fixed query $Q$, which instructs the model to identify objects, layouts, and materials, as illustrated in the Physics-Aware Vision-Language Priors block of Fig.~\ref{fig:2}. Formally, this process can be expressed as:
\begin{equation}
    \mathbf{T}_{Phys} = \text{VLM}(\mathbf{I}_{RGB}, Q).
\end{equation}
The generated $\mathbf{T}_{Phys}$ is then encoded by a pre-trained text encoder $\mathcal{T}(\cdot)$ (i.e., BERT~\cite{devlin2019bert}) to obtain semantic feature $\mathbf{F}_{Phys} \in \mathbb{R}^{M}$, as follows:
\begin{equation}
    \mathbf{F}_{Phys} = \mathcal{T}(\mathbf{T}_{Phys}).
\end{equation}
which provides physics-aware cues about material-dependent absorption and reflection, as well as object-level layout, thereby complementing the geometric features $\mathbf{F}_{RGB}$ and $\mathbf{F}_{Depth}$ in acoustic environment modeling.

\subsection{Binaural Audio Synthesis with Physics-Aware Feature Representation}
\noindent\textbf{\textit{Physics-Aware Feature Fusion:}} 
To obtain a unified representation of the acoustic environment, we propose an acoustic feature fusion adapter $\mathcal{A}(\cdot)$ that integrates geometric and semantic features. Specifically, the adapter fuses the multi-view geometric features $\mathbf{F}_{RGB}$ and $\mathbf{F}_{Depth}$ with the physics-aware semantic features $\mathbf{F}_{Phys}$:
\begin{equation}
    \mathbf{F}_{AFF} = \mathcal{A}(\mathbf{F}_{RGB}, \mathbf{F}_{Depth}, \mathbf{F}_{Phys}),
\end{equation}
where $\mathbf{F}_{AFF}$ denotes the fused physics-aware feature representation. 
Concretely, we concatenate the geometric features $\mathbf{F}_{RGB}$ and $\mathbf{F}_{Depth}$ and feed them into a multi-layer perceptron (MLP) to extract a geometric embedding, while the semantic features $\mathbf{F}_{Phys}$ are fed into another MLP to extract a semantic embedding. The outputs of these two MLPs are then added together to obtain the final fused feature $\mathbf{F}_{AFF}$.
This unified embedding jointly encodes room size, shape, object layout, and material properties, providing a physics-aware representation of the acoustic environment.

\noindent\textbf{\textit{Binaural Audio Generation:}} 
Finally, we employ the binaural audio generator $\mathcal{B}(\cdot)$ from AV-NeRF~\cite{liang2023av}, conditioned on the mono audio input $\mathbf{a}_{mono}$, the listener position $\mathbf{p}_L$, and the fused representation $\mathbf{F}_{AFF}$. The binaural signals are generated as:
\begin{equation}
    \mathbf{a}_{bi} = \mathcal{B}\left(\mathbf{a}_{mono} \mid \mathbf{F}_{AFF}, \mathbf{p}_L\right),
\end{equation}
where $\mathbf{a}_{bi}=\{\mathbf{a}_l,\mathbf{a}_r\}$ denotes the synthesized left and right audio channels. This design enables Phys-NVAS to generate binaural audio that reflects both geometric and semantic acoustic cues, leading to improved spatial realism and physical consistency.

\begin{table*}[htbp]
\centering
\caption{Performance comparison in terms of MAG and ENV. Lower MAG/ENV is better.}
{
\renewcommand{\arraystretch}{0.93}
\small
\begin{tabular}{l|cc|cc|cc|cc|cc|cc}
\toprule
\multirow{2}{*}{Methods} & \multicolumn{2}{c|}{Modality} & \multicolumn{2}{c|}{Office $\downarrow$} & \multicolumn{2}{c|}{House $\downarrow$} & \multicolumn{2}{c|}{Apartment $\downarrow$} & \multicolumn{2}{c|}{Outdoors $\downarrow$} & \multicolumn{2}{c}{Overall $\downarrow$} \\
& Audio & Visual & MAG & ENV & MAG & ENV & MAG & ENV & MAG & ENV & MAG & ENV \\
\midrule
Mono-Mono & \ding{51} & \ding{55} & 9.269 & 0.411 & 11.889 & 0.424 & 15.120 & 0.474 & 13.957 & 0.470 & 12.559 & 0.445 \\
Mono-Energy & \ding{51} & \ding{55} & 1.536 & 0.142 & 4.307 & 0.180 & 3.911 & 0.192 & 1.634 & 0.127 & 2.847 & 0.160 \\
Stereo-Energy & \ding{51} & \ding{55} & 1.511 & 0.139 & 4.301 & 0.180 & 3.895 & 0.191 & 1.612 & 0.124 & 2.830 & 0.159 \\
\midrule
INRAS~\cite{su2022inras} & \ding{51} & \ding{55} & 1.405 & 0.141 & 3.511 & 0.182 & 3.421 & 0.201 & 1.502 & 0.130 & 2.460 & 0.164 \\
NAF~\cite{luo2022learning} & \ding{51} & \ding{55} & 1.244 & 0.137 & 3.259 & 0.178 & 3.345 & 0.193 & 1.284 & 0.121 & 2.283 & 0.157 \\
\midrule
ViGAS~\cite{chen2023novel} & \ding{51} & \ding{51} & 1.049 & 0.132 & 2.502 & 0.161 & 2.600 & 0.187 & 1.169 & 0.121 & 1.830 & 0.150 \\
AV-NeRF~\cite{liang2023av} & \ding{51} & \ding{51} & 0.930 & 0.129 & 2.009 & 0.155 & 2.230 & 0.184 & 0.845 & 0.111 & 1.504 & 0.145 \\
\rowcolor{gray!20} \textbf{Phys-NVAS} & \ding{51} & \ding{51} & \textbf{0.856} & \textbf{0.126} & \textbf{1.984} & \textbf{0.154} & \textbf{2.098} & \textbf{0.180} & \textbf{0.787} & \textbf{0.109} & \textbf{1.431} & \textbf{0.142} \\
\bottomrule
\end{tabular}
}
\label{tab:RWAVS dataset}
\vspace{-10pt}
\end{table*}

\section{Experiments and Results}

\subsection{Experiment Setup}

\noindent \textbf{\textit{Dataset:}} 
We evaluate our method on the RWAVS dataset~\cite{liang2023av}, which provides multimodal samples including camera poses, images, and high-quality binaural audio. 
The dataset covers four scenes (office, house, apartment, and outdoor), with recordings of 10–25 minutes sampled at 1 fps. 
Each frame is paired with one-second binaural and source audio, forming a complete data sample. 
Following AV-NeRF~\cite{liang2023av}, we adopt the official 80/20 split, resulting in 9850 training and 2469 validation samples after pre-processing.

\noindent \textbf{\textit{Evaluation Metrics:}} 
We follow AV-NeRF and report two widely used metrics for spatial audio, i.e., magnitude distance (MAG)~\cite{xu2021visually}, which measures spectral amplitude differences, and envelope distance (ENV)~\cite{morgado2018self}, which measures envelope structure differences. Lower scores indicate better perceptual alignment with the ground truth.

\noindent \textbf{\textit{Baseline Methods:}} 
To validate the effectiveness of Phys-NVAS, we compare it with both signal-based and learning-based methods. 
The signal-based methods include \textit{Mono-Mono} (duplicating mono to both channels), \textit{Mono-Energy} (scaling mono by average energy), and \textit{Stereo-Energy} (constructing stereo from known energy priors). 
The learning-based methods are INRAS~\cite{su2022inras}, NAF~\cite{luo2022learning}, ViGAS~\cite{chen2023novel}, and AV-NeRF~\cite{liang2023av}. 
As Phys-NVAS is the first to explicitly incorporate physics-aware vision-language priors, our experiments emphasize validating its effectiveness rather than exhaustively benchmarking all AV-NeRF variants. 
We adopt the same binaural generator as AV-NeRF for direct comparison, ensuring fairness and representativeness.

\subsection{Performance Comparison}
Table~\ref{tab:RWAVS dataset} reports the results across all environments. Our proposed Phys-NVAS consistently outperforms the compared baseline methods with the lowest MAG and ENV scores. Compared with AV-NeRF~\cite{liang2023av}, Phys-NVAS gains stem from multi-view 3D acoustic modeling, which improves spatial awareness of direct sound and early reflections, and physics-aware vision–language priors, which capture object layout and material properties shaping reverberation and fine acoustic details. These complementary pieces of information provide physically grounded enhancements, validating the effectiveness of our Phys-NVAS with physics-aware feature representation in generating more realistic spatial audio\footnote{Demo examples are available at: \href{https://physnvas.github.io/}{https://physnvas.github.io/}.}.

\begin{figure}
    \centering
    \includegraphics[width=.95\linewidth]{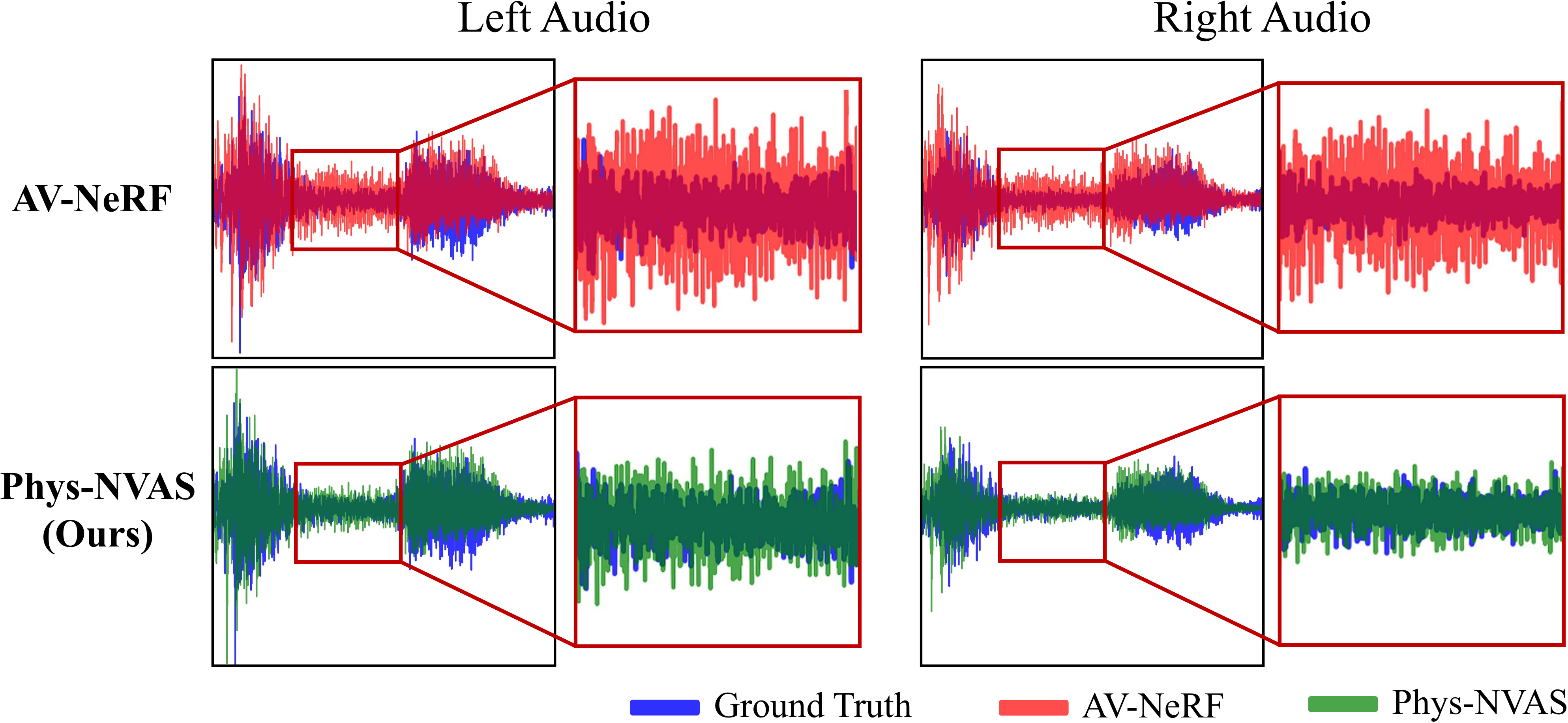}
    \vspace{-.25cm}
    \caption{Comparison of reconstructed binaural waveforms at a target listener position using AV-NeRF and our Phys-NVAS.}
    \label{fig:3}
   \vspace{-.33cm}
\end{figure}

\subsection{Visualization Analysis} 
Fig.~\ref{fig:3} compares binaural waveforms generated by AV-NeRF and the proposed Phys-NVAS with the ground truth. Phys-NVAS more closely follows the reference in both energy envelope and temporal dynamics, producing peaks with consistent amplitude and timing while exhibiting lower background fluctuations. In addition, it better preserves interaural differences, with channel energy asymmetry more closely aligned to the ground truth. These results indicate that Phys-NVAS captures scene acoustic characteristics more accurately and provides perceptually more reliable spatial cues for binaural audio generation.

\subsection{Ablation Study}
We evaluate the contribution of three feature sources, where RGB, DEP, and SEM denote $\mathbf{F}_{RGB}$ (multi-view appearance feature from 3DGS), $\mathbf{F}_{Depth}$ (depth-based structural feature), and $\mathbf{F}_{Phys}$ (semantic prior from the VLM), respectively. The baseline (first row in Table~\ref{tab:ablation 1}) uses only the binaural audio generator from AV-NeRF, conditioned on the mono audio input and listener position, without incorporating any additional priors.
As shown in Table~\ref{tab:ablation 1}, each feature individually improves performance, combining any two yields further gains, and using all three achieves the best results, confirming that geometric (RGB, DEP) and semantic (SEM) cues are complementary for accurate spatial audio synthesis.

\begin{table}[t]
\centering
\vspace{-.3cm}
\caption{Ablation study of feature components. 
}
\setlength{\tabcolsep}{8pt}
{
\renewcommand{\arraystretch}{0.93}
\small
\begin{tabular}{ccc|cc}
\toprule
\multicolumn{3}{c|}{\textbf{Feature Sources}} & \multicolumn{2}{c}{\textbf{Overall $\downarrow$}} \\
RGB & DEP & SEM & MAG & ENV \\
\midrule
          &     &     & 1.638 & 0.146 \\
\checkmark &     &     & 1.463 & 0.142 \\
          & \checkmark &     & 1.476 & 0.143 \\
          &     & \checkmark & 1.602 & 0.145 \\
\midrule
\checkmark & \checkmark &     & 1.448 & 0.143 \\
\checkmark &     & \checkmark & 1.463 & 0.143 \\
          & \checkmark & \checkmark & 1.468 & 0.143 \\
\midrule
\rowcolor{gray!20} \checkmark & \checkmark & \checkmark & \textbf{1.431} & \textbf{0.142} \\
\bottomrule
\end{tabular}
}
\label{tab:ablation 1}
\vspace{-.3cm}
\end{table}

\section{Conclusion}
This paper presented a physics-aware framework for novel-view acoustic synthesis that integrates multi-view spatial perception and vision–language priors to jointly model scene geometry, object layout, and material properties. Unlike prior audio-only or single-view visual methods, Phys-NVAS provides a unified representation that captures both geometric and physics-aware semantic cues critical for realistic acoustics. Experiments on the RWAVS dataset show consistent gains across environments, validating the effectiveness of Phys-NVAS and the complementarity of geometric and semantic cues.

\section{Acknowledgement}
This work was supported by the Fundamental Research Funds for the Central Universities (Grant No. 3072025YY0601).

\bibliographystyle{IEEEtran}
\bibliography{refs}

\end{document}